\documentclass[dvips,longauth,traditabstract]{aa}
\usepackage[dvips]{graphicx}
\usepackage{txfonts}
\usepackage{natbib}
\usepackage{longtable}
\usepackage{rotating}
\usepackage{pdflscape}
\usepackage{multirow}

\def\nhtwopeakb{1.3$\pm$ 0.1 $  \times10^{22}$}
\def\nhtwopeakl{3.6$\pm$ 0.1 $\times10^{22}$}
\def\nhtwosedb{1.0$\times10^{22}$}

\def\tpeakb{9.3$\pm0.5$}
\def\tpeakl{9.8$\pm 0.5$}
\def\tsedpeakb{11.3}
\def\tsedpeakl{11.6}

\def\tsedoutb{16.5$\pm$2}

\def\ximaxb{7.0$\pm$0.1}

\def\msol{$M_{\odot}$}
\def\lsol{$L_{\odot}$}

\def\overdensityb{17} 

\def\micron{$\mu$m}
\def\Td{$T_{\rm d}$}
\def\nhtwo{$N_{\rm H_2}$}
\def\av{$A_{\rm V}$}
\def\rv{$R_{\rm V}$}
\def\Rflat{$R_{\rm flat}$}
\def\kms{km~s$^{-1}$}
\def\csb{0.17}

\begin{document}

\title{Reconstructing the density and temperature structure of prestellar cores from \emph{Herschel} data: A case study for B68 and L1689B}

\author{A. Roy\inst{\ref{inst1}}
  \and Ph. Andr\'{e}\inst{\ref{inst1}}
  \and P. Palmeirim\inst{\ref{inst1}}
  \and M. Attard\inst{\ref{inst1}}
  \and V. K\"{o}nyves\inst{\ref{inst1},\ref{inst-doris}}
  \and N. Schneider \inst{\ref{inst1},\ref{inst-nicola-s}}
  \and N. Peretto\inst{\ref{inst1},\ref{inst-cardif}}
  \and A. Men'shchikov\inst{\ref{inst1}}
  \and D. Ward-Thompson\inst{\ref{inst-jason}} 
  \and J. Kirk\inst{\ref{inst-jason}}
  \and M. Griffin \inst{\ref{inst-cardif}}
  \and K. Marsh \inst{\ref{inst-cardif}}
  \and A. Abergel \inst{\ref{inst-doris}}
  \and D. Arzoumanian \inst{\ref{inst-doris}}
  \and M. Benedettini \inst{\ref{inst-rgyl}}
  \and T. Hill \inst{\ref{inst1},\ref{inst-tracey}}
  \and F. Motte \inst{\ref{inst1}}
  \and Q. Nguyen Luong\inst{\ref{inst-cita}}
  \and S. Pezzuto \inst{\ref{inst-rgyl}}
  \and A. Rivera-Ingraham\inst{\ref{inst-alana},\ref{inst-alana1}}
  \and H. Roussel \inst{\ref{inst-iap}}
  \and K.~L.~J. Rygl \inst{\ref{inst-rgyl}}
  \and L. Spinoglio \inst{\ref{inst-rgyl}}
  \and D. Stamatellos\inst{\ref{inst-jason}}
  \and G. White \inst{\ref{inst-white1},\ref{inst-white2}}
 }

\institute{Laboratoire AIM, CEA/DSM-CNRS-Universit\'{e} Paris Diderot,
IRFU / Service d'Astrophysique, C.E. Saclay, Orme des
Merisiers, 91191 Gif-sur-Yvette, France \label{inst1} 
\and Institut d’Astrophysique Spatiale, CNRS/Universit\'{e} Paris-Sud 11, 91405 Orsay, France \label{inst-doris}
\and Universit\'{e} de Bordeaux, Laboratoire d’Astrophysique de Bordeaux, CNRS/INSU, UMR 5804, BP 89, 33271, Floirac Cedex, France \label{inst-nicola-s}
\and School of Physics \& Astronomy, Cardiff University, Cardiff, CF29, 3AA, UK \label{inst-cardif}
\and Jeremiah Horrocks Institute, University of Central Lancashire, PR1 2HE, UK\label{inst-jason}
\and INAF-Istituto di Astrofisica e Planetologia Spaziali, via Fosso del Cavaliere 100, I-00133 Rome, Italy \label{inst-rgyl}
\and Joint ALMA Observatory, Alonso de \'{C}ordova 3107, Vitacura, Santiago, Chile \label{inst-tracey}
\and Canadian Institute for Theoretical Astrophysics, University of Toronto, 60 St. George Street, Toronto, ON M5S 3H8, Canada\label{inst-cita}
\and Universit\'{e} de Toulouse; UPS-OMP; IRAP;  Toulouse, France \label{inst-alana} 
\and CNRS; IRAP; 9 Av. colonel Roche, BP 44346, F-31028 Toulouse cedex 4,  France\label{inst-alana1}
\and Institut d’Astrophysique de Paris, UMR7095 CNRS, Universit\'{e} Pierre \& Marie Curie, 98 bis Boulevard Arago, F-75014 Paris, France \label{inst-iap}
\and Department of Physics and Astronomy, The Open University, Walton Hall Milton Keynes, MK7 6AA, United Kingdom\label{inst-white1}
\and RAL Space, STFC Rutherford Appleton Laboratory, Chilton Didcot, Oxfordshire OX11 0QX, United Kingdom\label{inst-white2}
 email: Arabindo.Roy@cea.fr \\
philippe.andre@cea.fr}

\titlerunning{Reconstructing density and temperature profiles} 
\abstract{
  Utilizing multi-wavelength dust emission maps acquired with
  \emph{Herschel}, we reconstruct  local volume density and dust
  temperature profiles for the prestellar cores B68 and L1689B using inverse-Abel
  transform based technique.  We present intrinsic radial 
  dust temperature profiles of starless 
  cores directly from dust continuum emission maps
  disentangling the effect of temperature variations along the line of
  sight which was previously limited to the radiative transfer
  calculations. The reconstructed dust temperature profiles show a
  significant drop in core center, a flat inner part, and a rising
  outward trend until the background cloud temperature is reached.
  The central beam-averaged dust temperatures obtained for B68 and L1689B are
  \tpeakb\ K and \tpeakl\ K, respectively, which are lower than the
  temperatures of \tsedpeakb\ K and \tsedpeakl\ K obtained from direct
  SED fitting. The best mass estimates derived by integrating the
  volume density profiles of B68 and L1689B are 1.6~\msol\ and
  11~\msol, respectively. 
  Comparing our results for B68 with the
  near-infrared extinction studies, we find that
  the dust opacity law adopted by the HGBS project,
  $\kappa_{\lambda}=0.1\times\left(\frac{\lambda}{300~\rm \mu
    m}\right)^{-2}$ cm$^{2}$ g$^{-1}$, agrees to within 50\% with the
  dust extinction constraints.  }

\maketitle

\section{Introduction}
Recent submillimeter observations with the \emph{Herschel} Space
Observatory \citep{pilbratt2010} and particularly the results obtained
as part of the $Herschel$ Gould Belt Survey (HGBS;
\citealp{andre2010}) have significantly improved our global
understanding of the early stages of low-mass star formation.  It is
now becoming clear that the formation of prestellar cores is
intimately related to the ubiquitous filamentary structure
present in the cold interstellar medium (ISM) \citep{arzoumanian2011}.
One of the main objectives of the HGBS is to measure the prestellar
core mass function (CMF) in 
nearby cloud complexes and to clarify 
the relationship between the CMF and the stellar
initial mass function (IMF) on one hand and the link 
with the structure of the ISM on the other hand  
\citep[cf.][for preliminary results]{konyves2010, andre2010}.  
  
An accurate determination of the
prestellar CMF requires reliable estimates of core masses.  In the
context of the HGBS project, core masses are derived from dust
continuum emission maps obtained with \emph{Herschel} between
160~\micron\ and 500~\micron .  Dust emission is almost always
optically thin at these wavelengths and can thus act as a surrogate
tracer of the total (gas $+$ dust) mass along the line of sight (LOS).
This requires an assumption about the dust opacity in the
submillimeter regime and reliable estimates of the dust temperature,
$T_{\rm d}$.  \emph{Herschel} multi-wavelength data can be used to
estimate $<$\Td$>_{\rm LOS}$ through single-temperature greybody
fits to the observed spectral energy distributions (SEDs)
\citep[cf.][]{konyves2010}.  A complication, however, is that such
fits only provide the average temperature along the LOS and do not
account for temperature gradients within the target sources.  This is
potentially a very significant problem when studying self-gravitating
starless (or protostellar) cores with stratified density structures
heated by an external or internal radiation field.  The central
temperatures of cold, starless cores, and sometimes their mass-averaged
temperatures, can be overestimated due to the relatively strong 
emission from the screen of warmer dust in the outer
layers of the cores, biasing core mass estimates to lower values \citep[cf.][]{malinen2011}.
Likewise, temperature variations along the LOS may hamper the
derivation of reliable density profiles for prestellar cores (e.g., \citealp{kirk2005,ysard2012}) 
 using
submillimeter emission maps (e.g., \citealp{kirk2005,ysard2012}).  Radiative
transfer calculations (e.g., \citealp{evans2001,stamatellos2007}) have
been performed to predict the dust temperature profiles of starless
cores whose outer surfaces are exposed to heating by the local
interstellar radiation field (ISRF).  These calculations generally
find a significant drop in $T_{\rm d}$ at the center of starless
cores, but the actual magnitude of this temperature drop remained
poorly constrained observationally before the advent of
\emph{Herschel} (see \citealp{ward-Thompson2002} for early results 
with  \emph{ISO}, however).

Here, we use \emph{Herschel} observations to quantify the dust
temperature gradient within two well-studied starless cores, B68 and
L1689B.  We introduce a simple yet powerful inversion technique based
on the Abel integral transform to simultaneously reconstruct the 3D
density and dust temperature profiles of dense cores using as inputs
\emph{Herschel} maps in four bands  between 160~\micron\ and
500~\micron\  (also see \citealp{marsh2014}  for an independent approach of
solving similar problem).
The basic goal of the present paper is to demonstrate the performance
of the Abel inversion technique.  We apply our algorithm to B68 and
L1689B and compare the results with previous infrared
extinction/absorption studies, which allows us to validate the dust
opacity assumption made in earlier HGBS papers.  We also test our
technique on synthetic core models with known density and temperature
profiles.  While the method assumes spherically symmetric cores, we
show that it yields satisfactory results for prolate or oblate
ellipsoidal cores with realistic aspect ratios $\la 2 $.

\section{\emph{Herschel} Observations of B68 and L1689B}\label{sec:obs}

As a part of the HGBS key project, two fields in the Pipe and Ophiuchus 
molecular cloud complexes containing B68 and L1689B, of areas 
$\sim 1\rlap{.}\degr 5 \times 1\rlap{.}\degr5$ and $\sim 3\rlap{.}\degr0 \times 3\rlap{.}\degr5$
%
and observed for an integration time of 2.6 and 10.2 hours, respectively.
These target fields were mapped\footnote{A more detailed description about the observations and
data reductions are also available in the HGBS website: http://gouldbelt-herschel.cea.fr/archives.} 
in two orthogonal scan directions 
at a scanning speed of 60\arcsec\ s$^{-1}$ in parallel mode,
 acquiring data  simultaneously in five bands with the SPIRE \citep{griffin2010} and PACS
\citep{poglitsch2010} bolometer cameras.  
The data were reduced using
HIPE version 7.0.  The SPIRE data were processed with modified
pipeline scripts.  Observations during the turnaround of the telescope
were included, and a destriper module with a zero-order polynomial
baseline was applied. The default 'na\"{\i}ve' mapper was used to
produce the final map.  
For the PACS data, we applied the standard HIPE data reduction pipeline up to level 1, 
with improved calibration. Further processing of the data, such as subtraction of  
(thermal and non-thermal) low-frequency noise and map projection was performed 
with Scanamorphos v11 \citep{roussel2012}. Note that the  Scanamorphos  map-maker avoids any
high-pass filtering  which is crucial for preserving extended emission.

\begin{figure}
  \centering
  \resizebox{\hsize}{!}{\includegraphics[angle=0]{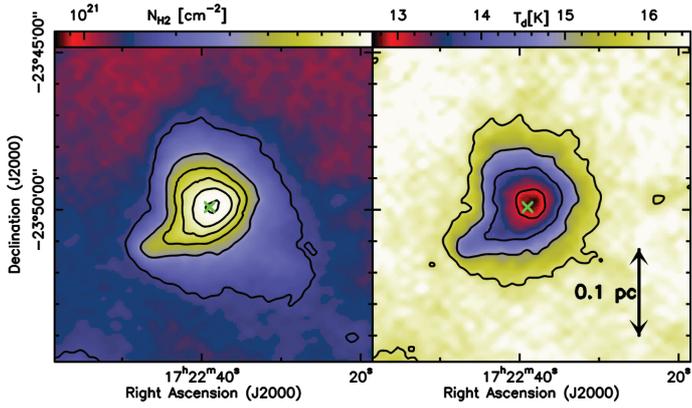}}
  \caption{Column density map (left) and LOS dust temperature map
    (right) of B68 derived by fitting modified blackbody SEDs to the
    \emph{Herschel} data between 160~\micron\ and 500~\micron\ on a
    pixel-by-pixel basis (see Sect.~\ref{sec:sedfit} for details).
    The column density contours go from $1.6 \times 10^{21}$ to $9.6
    \times 10^{22}$ by steps of $1.6 \times 10^{21}$ in units of H$_2$
    molecules per cm$^2$ (left), and the temperature contours are 16, 15, 14, 13~K (right).  
    The cross symbol shows the center of the core obtained by
    fitting a 2D-Gaussian to the column density map.}
\label{fig:B68-nhT}
\end{figure}

\begin{figure}
  \centering
  \resizebox{\hsize}{!}{\includegraphics{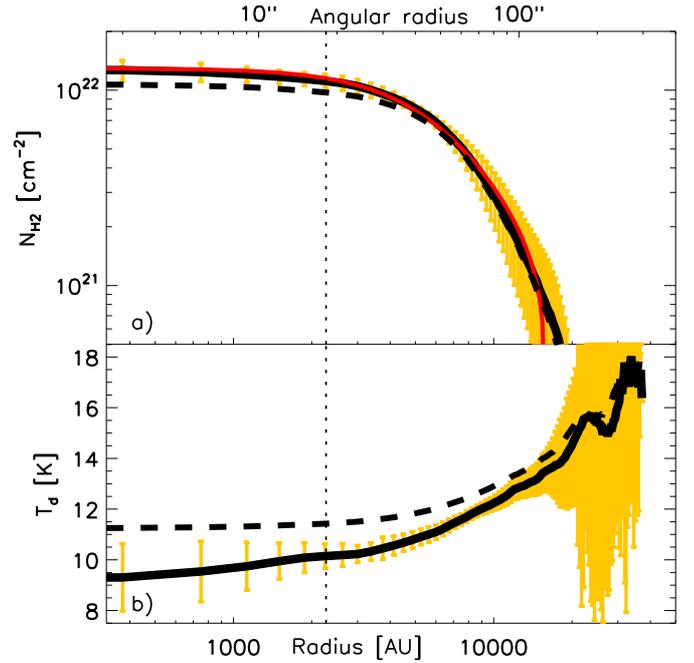}}
  \caption{Column density ({\bf a}) and dust temperature ({\bf b})
    profiles of B68 obtained at 500~\micron\ resolution by applying
    the Abel inversion method to the circularly-averaged intensity
    profiles observed with \emph{Herschel} between 160~\micron\ and
    500~\micron.  {\bf a}) Comparison between the column density
    profiles derived from the Abel reconstruction (solid black line)
    and from LOS-averaged SED fitting (thick dashed line).
    The red curve
    shows the best-fit Bonnor-Ebert model to the Abel-inverted profile
    (see Table~\ref{tab:bonnor} for parameters). The vertical  dotted line
    represents the half power beam radius of 36\farcs3/2 (effective 500~\micron\ resolution).
    {\bf b}) Comparison of the Abel-inverted (solid line) and
    LOS-averaged SED (dashed curve) temperature profiles.
    The error bars represent the standard deviation
    of \nhtwo(r) and \Td(r) values obtained from independent profile
    reconstructions along sixteen angular directions.  }
  \label{fig:B68-profile}
\end{figure}

\section{Description of the Abel inversion method}\label{sec:abel}

Consider a spherically symmetric core with radial density profile,
$\rho(r)$, embedded in a uniform background and isotropic  ISRF.  Assuming optically thin
dust emission, the specific intensity $I_\nu(p)$ of the core when
observed at impact parameter, $p$, may be expressed as:

\begin{equation}
I_\nu (p) = 2\, \int_{p}^{+\infty}\, \rho (r)\, B_\nu\left[T_{\rm
    d}(r)\right]\, \kappa_\nu \frac{r\,dr}{\sqrt{r^2-p^2}}+ I_{\nu,\rm
  bg} +I_{\nu, \rm N},
\label{eq:a}
\end{equation}
where $I_{\nu,\rm bg}$ and $I_{\nu, \rm N}$ represent the 
background emission and instrumental noise, respectively,
$B_\nu[T_{\rm d}(r)]$ is the Planck function for
the dust
temperature $T_{\rm d}(r)$ at radius $r$ from  core center, and
$\kappa_{\nu}$ is the frequency-dependent dust opacity\footnote{
Note the dust-to-gas fraction of 1\%
is implicitly included  in our definition of the dust opacity (see also Sect. \ref{sec:sedfit})
so that $\rho(r)$ represents the
radial gas density of the object.}, here assumed to
be uniform throughout the core. 
Given the
symmetry of the problem, we can use the inverse Abel transform
\citep[e.g.][]{bracewell1986} to obtain the integrand of
Eq.~(\ref{eq:a}) at each observed frequency $\nu$:
\begin{equation}
\rho (r)\, B_\nu\left[T_{\rm d}(r)\right]\, \kappa_\nu = 
-\frac{1}{\pi}\, \int_{r}^{+\infty}\, \frac{dI_\nu}{dp}
\frac{dp}{\sqrt{p^2-r^2}}.
\label{eq:b}
\end{equation}
From Eq.~\ref{eq:b}, we see that the physical parameters of interest, 
$\rho(r)$ and $T_{\rm d}(r)$, only depend on 
the first derivatives of the radial intensity profiles.
With a pre-defined assumption about the dust opacity law
$\kappa_{\nu}$, one may thus estimate $T_{\rm d}(r)$ at each radius
$r$ by fitting a single-temperature modified blackbody to the SED
obtained from evaluating the right-hand side of Eq.~(\ref{eq:b}) at
each observed band between 160~\micron\ and 500~\micron\ (see
Sect.~\ref{sec:sedfit} for further details).  The density profile,
$\rho(r)$, can be derived simultaneously from the normalization of the fit
at each radius.

For simplicity, Eq.~(\ref{eq:a})  above neglects the
convolution with the telescope beam.  
However, simulations confirm that beam smearing has little effect in the case of well-resolved
cores (see Appendix ~\ref{appen:sim}).  In the case of starless cores
such as B68 and L1689B with a flat inner density profile inside a
radius \Rflat, or angular radius $\theta_{\rm flat} $, we find that
the beam effect can be parameterized by the ratio $\theta_{\rm
flat}/HPBW$ (where $HPBW$ is the half-power beam width), and that
for $\theta_{\rm flat}/HPBW \ga 1$, as is the case for B68 and L1689B,
the reconstructed column density profile agrees with the intrinsic 
profile within 20\% (in the absence of noise).  
More generally, simulations indicate that, in the absence of noise, the reconstructed temperature 
and column density profiles essentially coincide with the corresponding intrinsic profiles 
{\it convolved} to the effective beam resolution (see Appendix ~\ref{appen:sim}). 
Importantly, our Abel
inversion technique does not depend on the subtraction of a flat
background level as the derivative of a constant background\footnote{
If the background is not flat, the input data can be pre-processed
by subtracting e.g., a linear plane.}
does not contribute to the integral of Eq.~(\ref{eq:b}).  The reconstruction 
is, however, quite sensitive to noise fluctuations in the outer parts
of the core.

\begin{figure}

\centering
\resizebox{\hsize}{!}{\includegraphics[angle=0]{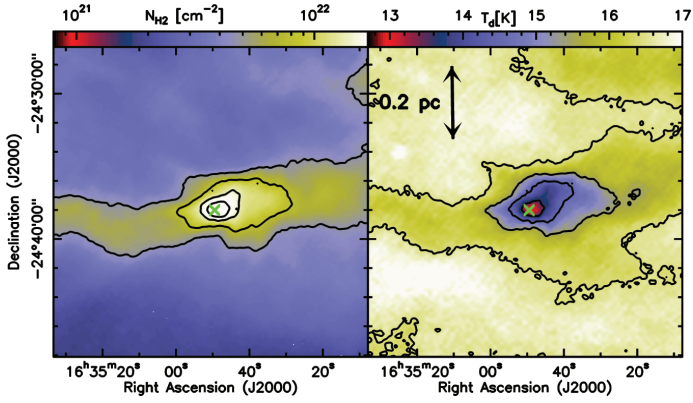}}
\caption{Same as Fig.~\ref{fig:B68-nhT} but for L1689B.
The column density contours are 6$\times 10^{21}$, 1$\times 10^{22}$,
1.4$\times 10^{22}$, 1.8$\times 10^{22}$, and 2.6$\times 10^{22}$
H$_2$  cm$^{-2}$ (left),
and the temperature contours are  16.5, 15.5, 14.5, and 13.5~K (right).
The column density image shows that L1689B is embedded inside a
filamentary structure. }
\label{fig:L1689B-nhT}
\end{figure}

\begin{figure}
\centering
\resizebox{\hsize}{!}{\includegraphics{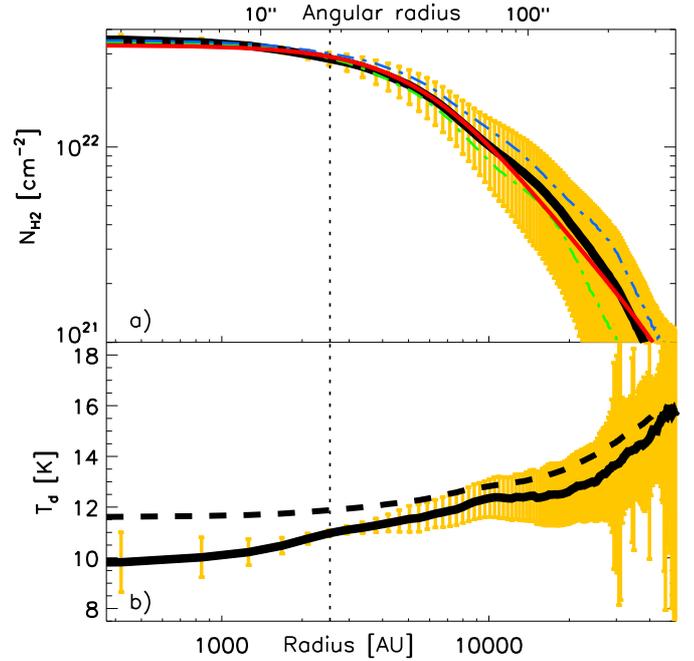}}
\caption{ {\bf a)} Column density profile of L1689B obtained at
  500~\micron\ resolution by applying the Abel reconstruction method
  to the circularly-averaged intensity profiles between
  160~\micron\ and 500~\micron\ (black solid curve).  The blue and
  green dash-dotted profiles show the column density profiles obtained
  from reconstructing the intensity profiles observed along EW and NS
  sectors, respectively.  The overplotted red solid line shows the
  best-fit Bonnor-Ebert model (see Table~\ref{tab:bonnor} for
  parameters).  The vertical  dotted line is same as in Fig.~\ref{fig:B68-profile}.
  {\bf b)} Reconstructed dust
  temperature profile of L1689B (solid curve) compared with the
  LOS-averaged temperature profile derived from simple SED fitting
  (thick dashed curve).  }
\label{fig:L1689B-profiles}
\end{figure}

\subsection{SED fitting and profile reconstruction }\label{sec:sedfit}

Column density and corresponding LOS-averaged dust temperature maps are shown in
Fig.~\ref{fig:B68-nhT} and Fig.~\ref{fig:L1689B-nhT} for B68 and
L1689B, respectively. They were obtained by fitting modified blackbody
functions to the \emph{Herschel} SEDs longward of 160~\micron\ on a
pixel-by-pixel basis as in earlier HGBS papers
(\citealp{konyves2010}; see also \citealp{hill2011} for HOBYS). 
While fitting the SEDs, we weighted 
each data point by the corresponding calibration errors at SPIRE ($\sim 10$\%)
 and PACS ($\sim 15$\%) wavelengths. 
Appropriate zero-level offsets were added to each image (see
Table~\ref{tab:zerolevel}), obtained by correlating the
\emph{Herschel} data with the \emph{Planck} and \emph{IRAS} data of
the same fields \citep[see][]{bernard2010}.  The same dust opacity law
as in earlier HGBS   (see also \citealp{motte2010} for the HOBYS key program) papers is 
adopted in the present paper (similar to
\citealp{hildebrand1983}):
$\kappa_{\lambda}=0.1\times\left(\frac{\lambda}{300 \mu{\rm
    m}}\right)^{-\beta} $ cm$^{2}$ per g (of gas $+$ dust), with a
dust emissivity index of $\beta = 2$. Our dust opacity value at the normalizing wavelength is also
close to the \cite{ossenkopf1994} opacity model for dust grains with thin ice mantles.
A mean molecular weight
$\mu_{\rm H_2} = 2.8$ is assumed\footnote{Note that this differs
  from the first HGBS papers
  \citep[e.g.][]{andre2010,konyves2010,arzoumanian2011} where $\mu =
  2.33$ was assumed and column density was expressed in units of mean
  free particles per cm$^2$.} to express column density in units of
${\rm H_2}$ molecules per cm$^2$. Using our adopted  dust opacity law,  
we find that even the central LOSs for B68 and L1689B  with \nhtwo\ $\leq$ 5$\times10^{22}$ cm$^{-2}$ 
has small optical depths $\lesssim$ 0.08 at 160~\micron, confirming that the core emission is
optically thin longward of 160~\micron.

The same assumptions have been adopted for the modified blackbody fits
required at each radius by the Abel inversion technique.  A
single-temperature description of the SEDs is suitable in this case
because we are sampling local density and dust temperature at a given
radius $r$.  
Prior to SED fitting, the
\emph{Herschel} data are convolved to a common resolution of either 
36\farcs3 (i.e., HPBW resolution of SPIRE at 500~\micron ) 
or 24\farcs9\  (i.e., HPBW resolution of SPIRE at 350~\micron , when the 500~\micron\  
data are not used).
At each wavelength, a circularly-averaged intensity
profile about core center is first derived from the
\emph{Herschel} data, and the derivative of this average profile is
then numerically evaluated and integrated over the kernel shown in the
right-hand side of Eq.~(\ref{eq:b}).  The central position of the core
is obtained from fitting a 2D-Gaussian model to the column density map
(see the cross symbols in Fig.~\ref{fig:B68-nhT} and
Fig.~\ref{fig:L1689B-nhT} for B68 and L1689B, respectively).  The
integration is performed up to an outer radius corresponding to $\sim
$~2--2.5 times the $FWHM$ {\it diameter} of the core as estimated from
the 2D-Gaussian fit to the column density map. In practice, the
integral on the right-hand side of Eq.~(\ref{eq:b}) converges
rapidly and does not depend much on the precise value of the integration
outer radius so long as it encompasses the entire core -- see
Appendix~\ref{appen:profile} and Fig.~\ref{fig:AA1}.

\subsection{Assessment of uncertainties}
The errors bars on the reconstructed profiles at each radius can be
estimated from the standard deviations of the density and temperature
profiles obtained by repeating the Abel reconstruction along different
angular directions around the source (i.e., averaging the data
separately over a series of angular sectors instead of
circularly-averaging the data).  In the case of the reconstructed
profiles of B68 and L1689B shown in Fig.~\ref{fig:B68-profile} and
Fig.~\ref{fig:L1689B-profiles}, sixteen equally-spaced angular sectors
were used to derive the error bars.  The errors we report on the best
estimates of the (column) density and temperature at core center
correspond to the standard error in the mean, $\sigma/\sqrt{n} $,
where $n$ is the number of independently measured sectors.

The uncertainty in the normalization of the dust opacity law directly
affects the column density estimates but not the temperature estimates, 
whereas the uncertainty in the
emissivity index $\beta$ influences both the dust temperature and the
column density estimates in an anti-correlated fashion.  When $\beta$
is varied from 2 to 1.5, for instance, the dust temperature increases by $\sim$10\% 
and the column density decreases by $\sim$40\%. 
The uncertainty in the central position of the
core also introduces errors on the derived parameters.
To assess the magnitude of this effect in the case of B68, we varied
the central position within a radius of 6\arcsec, and found that the
resulting central dust temperature and central column density had 
standard deviations of $\pm$ 0.3 K, and $\pm$0.04$\times 10^{22}$
cm$^{-2}$, respectively.  Besides, departures from spherical symmetry
due to, e.g., asymmetries in the density distribution (cf.  L1689B in
Fig.~\ref{fig:L1689B-nhT}) or an anisotropic background radiation
field \citep[cf.][]{nielbock2012} may break the symmetry of the
quantity $\rho(r) B_{\nu}(T(r))$ in Eqs.~(\ref{eq:a}) and (\ref{eq:b}), 
introducing
additional errors. The latter are in principle included in the error
bars estimated from the fluctuations of the radial profiles
reconstructed along various angular directions.  Moreover, we show in
Appendix~\ref{appen:sim} that cores with moderate departures from
spherical symmetry (such as prolate or oblate ellipsoidal cores with
aspect ratios $\la 2 $) can be reconstructed with reasonable accuracy.

For both B68 and L1689B, the net measurement errors  in 
the derived central $T_{\rm d}$
and \nhtwo values are estimated
to be $\pm$0.5 K and $\pm 0.1 \times 10^{22} $ cm$^{-2}$, respectively,
excluding the systematic uncertainties associated with our
assumptions on the dust opacity and the calibration errors. 
In Appendix~\ref{A:sph} we show
that the  calibration errors  lead to an additional
uncertainty of $\sim$12\%  on column density and $\sim$5\% on temperature
estimates.

\begin{center}
\begin{table}
\begin{minipage}{1\linewidth}
\caption{Planck offsets in MJy sr$^{-1}$}\label{tab:zerolevel}
\begin{tabular}{|c c c c c|}
\hline
Target & 160 $\mu$m       & 250  $\mu$m      & 350 $\mu$m       & 500 $\mu$m \\
\hline
B68    &  91.2            & 77.0             & 41.6             & 16.6 \\
L1689B &  89.3            & 136.3            & 64.4             & 26.6 \\
\hline
\end{tabular}
\end{minipage}
\end{table}
\end{center}

\section{Detailed results for B68 and L1689B}\label{sec:result}
\subsection{B68 core}\label{sec:B68}

B68 is a well studied isolated bok globule in the Pipe nebula cloud
complex \citep[e.g.][]{alves2001,nielbock2012}. We adopt a distance of
125 pc \citep{de-geus1989} for the present study.
Figure~\ref{fig:B68-profile} shows the reconstructed column density
and temperature profiles obtained from our \emph{Herschel}
observations of B68 with the Abel inversion method described in
Sect.~\ref{sec:abel}.

In Fig.~\ref{fig:B68-profile}b, the Abel-inverted radial dust
temperature profile (solid curve) is compared with the LOS-averaged
SED temperature profile (dashed curve). While both  temperature
profiles exhibit similar features, with a broad minimum around the
core center, a positive gradient outside the flat inner plateau of the
column density profile, and similar values (\Td $\sim$ \tsedoutb~K) at
large radii, the minimum Abel-reconstructed temperature at core center
is $\sim 2 $~K lower than the minimum SED temperature observed through
the central LOS.  Accordingly, the central column density derived with 
the Abel inversion method (\nhtwopeakb\ cm$^{-2}$) is 30\%
higher than the LOS-averaged column density of  $\sim $\nhtwosedb\ cm$^{-2}$ 
derived from standard SED fitting for the central LOS.  
The difference between the Abel-reconstructed and the LOS-averaged
column density 
becomes negligible in the outer parts of the
core.  This is indicative of stronger temperature variations
along the central LOS  compared to the outer LOSs.

Recently, \cite{nielbock2012} constrained the dust temperature and
volume density profiles of B68 using an iterative approach based on 3D
radiative transfer modeling of multi-wavelength dust continuum data
including \emph{Herschel} observations obtained as a part of the EPoS \citep{launhardt2013}
key project.  They employed a 3-dimensional 
grid of Plummer-like \citep{plummer1911} density profiles and obtained
initial guesses of the parameters from LOS-averaged SED fits.
Altogether their model was tuned with eight free parameters, yielding
a central dust temperature of \Td = 8.2$^{+2.1}_{-0.7}$~K.
The primary reason for the dispersion in the central dust temperature
in \cite{nielbock2012} is the uncertainty of a factor of $\sim 2$ 
on the dust opacity in the infrared regime.  Although we obtain a higher
central dust temperature, \Td=\tpeakb ~K, with our Abel-inversion
method, our results agree with the Nielbock et al. analysis within the
range of the quoted uncertainties.

Based on extinction measurements \citep{alves2001}, the column density profile of
B68 closely resembles that of a Bonnor-Ebert (BE) (e.g., \citealp{bonnor1956}) isothermal sphere
with a flattened inner region.  The overplotted red curve in
Fig.~\ref{fig:B68-profile} shows the best-fit BE model to the
reconstructed column density profile.  The best-fit BE parameters
along with the physical properties that follow directly from the fit
such as the radius of the flat inner plateau, $R_{\rm flat} \equiv
2\,\rm{c}_s/\sqrt{4\pi G \rho_{\rm c } } $, the density contrast,
$\rho_{\rm c}/\rho_{\rm s}$, and the external pressure, $P_{\rm ext}$,
are summarized in Table~\ref{tab:bonnor}.  They are consistent  with the BE
parameters found by \cite{alves2001}.

The  \nhtwo\ column density profile obtained by integrating the
Abel-inverted volume density profile is shown in
Fig.~\ref{fig:B68-profile}. This  can be directly compared with the
near-infrared extinction results \cite{alves2001} because extinction
traces material independently of temperature.  For this comparison, we
adopted a standard conversion factor, \nhtwo/\av\ of 9.4 $\times
10^{20}$ cm$^{-2}$~mag$^{-1}$ \citep{bohlin1978}, to translate column
density to equivalent visual extinction \av\ at low column densities
(i.e., \nhtwo\ $\la 6 \times 10^{21}\, {\rm cm}^{-2}$).  For higher
column densities, we used the conversion \nhtwo/\av =
6.9$\times10^{20}$ cm$^{-2}$~mag$^{-1}$ given by \cite{draine2003},
and later on adopted by \cite{evans2009}, consistent with an extinction
curve with a  total-to-selective extinction, \rv $ \equiv$ \av/$E(B-V) = 5.5$, 
appropriate in higher density regions.

Using the Bohlin  conversion factor, we obtain an \av\ of 3.2
mag at a radius of 10$^4$ AU, agreeing within $\sim$ 60\% with the
value of 5 mag reported by \cite{alves2001} at the same radius.
Furthermore, if we correct our estimate for the weak empirical trend
between submillimeter dust opacity and column density,
$\kappa_{\lambda} \propto \, $\nhtwo$^{0.28}$, inferred by
\cite{roy2013} in the regime $1 \la A_V \la 10$ and interpreted as
evidence of dust grain evolution, then we find a corrected column
density $ N^{\rm corr}_{\rm H_2} \approx 5 \times10^{21}\, {\rm
  cm}^{-2} $, equivalent to a corrected $A_V^{\rm corr} = 5.3$ mag, in
excellent (10\%) agreement with \cite{alves2001} at 10$^4$ AU.

Using the Draine conversion factor, we derive an \av\ of 19$\pm$2 mag
through the center of the B68 core compared to the \av\ of 30 mag
obtained from extinction by \cite{alves2001}, corresponding to a 60\%
agreement.  However, the effective angular resolution of the Alves et
al. extinction map was about $\sim$10\arcsec, approximately four times
higher than the beam resolution (36\farcs3) of SPIRE at 500~\micron.
For better comparison, we performed a similar Abel reconstruction
analysis at the beam resolution (24\farcs9) of the SPIRE
350-\micron\ observations, ignoring the 500~\micron\ data.  
In this case, the central Abel-reconstructed column density corresponds to
$A_V = 20 \pm 2$ mag, which agrees within 50\% with the results of \cite{alves2001}. 
(The central dust temperature derived at 350-\micron\  resolution coincides within 
the uncertainties with the 500-\micron\ resolution estimate.)

Likewise, the total mass of 1.6$\pm$0.07~\msol\ derived 
by integrating the Abel-reconstructed density profile of B68 within
the outer radius of 1.4$\times10^4$ AU agrees within 30\% with the mass of 2.1~\msol\ obtained by \cite{alves2001} from
extinction data.  For comparison, the mass derived from fitting a modified blackbody to the integrated flux densities 
is 1.4$\pm$0.05~\msol.
The latter does not account for temperature variations along the
LOS, and thus slightly underestimates the intrinsic total mass of the
core.  However, we stress that, for a moderate density core such as
B68, the global SED temperature (12.5$\pm$ 0.1 K here) is close to
the mass-averaged dust temperature and the SED mass agrees within
$\sim 15\% $ with the Abel-reconstructed mass.


 Using the above results on the density and temperature structure 
we can check the energy balance of the B68 core. Assuming optically 
thin submillimeter emission, a total output luminosity of 0.35$\pm$0.04~\lsol\ is 
obtained by integrating the quantity $4\pi\int\rho(r)B_{\nu}[T_{\rm d}(r)]\kappa_{\nu}d\nu$ 
over the volume of the core. 
A very similar output luminosity (0.41$\pm$0.05~\lsol) 
is estimated by integrating the observed SED over wavelengths. These output estimates 
should be compared to the input luminosity of $\sim$ 0.40~\lsol\ provided to the core 
by the local ISRF (\citealp{mathis1983}; with $G_0$ $\sim$1), calculated from the total ISRF flux 
density absorbed\footnote{The ISRF energy is mostly absorbed at short wavelengths (0.095~\micron\ to 100~\micron). 
In calculating
the absorbed energy we  adopted the dust absorption model of \cite{draine2003} with \rv=3.1.} 
by a spherical object with the same density profile and outer radius as B68 using Eq. 4 of \cite{lehtinen1998}.
Note that the above three luminosity values agree with one another.


\begin{center}
\addtolength{\tabcolsep}{-4pt}
\begin{table}
\begin{minipage}{\linewidth}
\caption{Best-fit parameters of Bonnor-Ebert core models for B68 and L1689B } \label{tab:bonnor}
\begin{tabular}{|cccccccc|}
\hline
Source & $\xi_{\rm max}$ & $\rho_{\rm c}/\rho_{\rm s}$\footnote{$\rho_{\rm c}$ and $\rho_{\rm s}$ denote central density and density at the outer surface of the BE sphere, respectively }   & c$_s$\footnote{ Isothermal sound speed }   & $P_{\rm ext}$ & $R_{\rm flat}$ \footnote{Flat inner radius, defined as $R_{\rm flat}  \equiv 2\,c_{\rm s}/\sqrt{4\pi G\rho_{\rm c} }$}&  $R_{\rm out}$ & $n_{\rm c}$  \\ \
profile       &                 &                         &   (\kms)  & (K cm$^{-3}$) &  (AU) &  (AU) & (cm$^{-3})$            \\
              &                 &                         &           &  $\times10^4$ & $\times10^3$ & $\times10^4$ & $\times10^4$ \\
\hline
B68    & \ximaxb                 &  \overdensityb              & \csb      & 3.9      &4.4   & 1.5    &   8.3 \\
L1689B-C &16.6$\pm$0.7             & 145                     & 0.27      & 2.9      & 4.5   & 3.7 &    20 \\
L1689B-NS &13.5$\pm$0.3            &88                        & 0.24     & 3.2      & 4.4    & 2.9 &    17\\
L1689B-EW &16.2$\pm$1.7            &137                  & 0.26      & 2.2     & 4.7   & 3.8 &   16.5\\
\hline
\end{tabular}
\end{minipage}
\end{table}
\end{center}

\subsection{L1689B}\label{sec:L1689B} 

Compared to B68, L1689B is a slightly denser and more centrally
condensed core \citep[e.g.][]{andre1996, bacmann2000}, located in the
Ophiuchus complex at a distance of $\sim 140$ pc.  The \emph{Herschel}
images reveal that it is embedded within a larger-scale filamentary
structure (see Fig.~\ref{fig:L1689B-nhT}).  Given the elongated
morphology observed in the plane of the sky and the mean apparent
aspect ratio $\sim 1.3$ of the core, the hypothesis of spherical
symmetry underlying  Eq.~(\ref{eq:a}) is not 
strictly verified.  

In order to assess the validity of our Abel-inversion scheme in this
case, we therefore performed test reconstructions for prolate and oblate
ellipsoidal model cores with similar aspect ratios.  These tests
suggest that small departures from spherical symmetry have little
impact on the reconstruction results (see Appendix~\ref{appen:sim}).  

The upper panel of Fig.~\ref{fig:L1689B-profiles} shows the results
of three distinct reconstructions of the column density profile of
L1689B, obtained by applying the Abel inversion method to a) the
circularly-averaged intensity profiles (thick solid curve), b) the
intensity profiles observed in sectors oriented East-West (EW), i.e.,
approximately along the apparent major axis of the core (blue
dash-dotted curve), and c) the intensity profiles observed in sectors
oriented North-South (NS), i.e., approximately along the apparent
minor axis of the core (green dash-dotted curve).  While the three
column density profiles agree with each other within the central
plateau region, the EW profile lies 30\% above the NS profile at large
radii $\ga 10^4$~AU, and the circularly-averaged profile is
intermediate between the other two.  For reference and comparison with
B68, a BE model was fitted to each of the three column density
profiles and the results of these fits are given in
Table~\ref{tab:bonnor} (for the sake of clarity, only the model fit to
the circularly-averaged intensity profiles is shown in
Fig.~\ref{fig:L1689B-profiles}, as a red solid curve).  All three
fits yield a consistent value for the radius of the flat inner
plateau, $R_{\rm flat} \sim 4500 \pm$100~AU (corresponding to $\sim
32\arcsec $).
The average reconstructed column density within \Rflat\ is
3.5$\pm$0.1$\, \times \,10^{22}$ cm$^{-2}$, in good ($\sim \, $30\%)
agreement with the H$_2$ column density averaged over the flat inner
part of the core of 4.5--4.7$\times10^{22}$ cm$^{-2}$ found by
\cite{bacmann2000} based on their ISOCAM mid-infrared absorption study
(see their Table 2).  

Our best estimate of the total core mass obtained from Abel
reconstructing the circularly-averaged intensity profiles is 11$\pm$2~\msol.  
The tests we performed for ellipsoidal synthetic cores
(cf. Appendix~\ref{appen:sim}) indicate that the relative error in
this mass introduced by the departure from spherical symmetry is less
than 4\% for an intrinsic aspect ratio of $\sim 1.3$.  Our mass
estimate is in excellent agreement with the total mass reported by
\cite{bacmann2000}.  The SED mass derived from fitting a modified
blackbody to the integrated flux densities 
is 7.8$\pm$0.2~\msol, i.e., about 30\% lower than our best mass estimate.
The corresponding global SED temperature of the core is 12.5$\pm$0.2 K.

The central dust temperature obtained for L1689B using the Abel
transform technique is \tpeakl\ K (see lower panel of
Fig.~\ref{fig:L1689B-profiles}).  This value is higher than the very
low central dust temperature of 7.5~K advocated by \cite{evans2001} 
based on radiative transfer calculations assuming a
standard ISRF ($G_0 = 1$).  At least part of this discrepancy can be
explained by the fact that the effective external radiation field for L1689B is
about one order of magnitude stronger than the standard ISRF, due to
the presence of early-type stars in the immediate vicinity of the
Ophiuchus cloud \citep[cf.][]{liseau1999}.

\section{Conclusions}\label{sec:discussion}

The results on B68 and L1689B discussed in Sect.~\ref{sec:result},
along with the tests on model cores presented in Appendix~B,
demonstrate that the Abel transform technique can successfully produce
meaningful radial density and dust temperature profiles for
spatially resolved prestellar cores.  The most rewarding result is the
reconstruction of a radial dust temperature profile directly from
\emph{Herschel} dust continuum imaging data, independently of any
radiative transfer model. The Abel inversion technique  
(Sect.~\ref{sec:abel}) is very general, insensitive to background
subtraction to first order, and can perform equally well for
externally-heated (isotropically) starless cores or internally-heated  protostellar
cores.  While in principle the technique assumes spherically symmetric
cores, tests performed on ellipsoidal cores suggest that satisfactory
results are obtained even when the hypothesis of spherical symmetry is
not strictly valid (such as in the L1689B case --
cf. Fig.~\ref{fig:L1689B-nhT}).  A similar Abel-transform scheme may 
also be employed to reconstruct the intrinsic beam-averaged density and
temperature profiles of (approximately) cylindrically symmetric
filaments from the observed radial intensity profiles averaged along
the filament main axes.

For both B68 and L1689B we find a characteristic dip in the dust
temperature profile, with minimum beam-averaged values of \tpeakb\ K and \tpeakl\ K
at core center, respectively. The temperature profile smoothly merges
with the background cloud temperature at the  outer core radii.

The Abel transform technique yields central beam-averaged H$_2$ 
densities of $7.5 \pm 0.5 \times 10^4\, {\rm cm}^{-3}$ and $2.0 \pm
0.1 \times 10^5\, {\rm cm}^{-3} $ for B68 and L1689B, respectively,
corresponding to central column densities of \nhtwopeakb\ cm$^{-2}$
and \nhtwopeakl\ cm$^{-2}$ after integration of the reconstructed
volume density profiles along the LOS.  These central column density
estimates are approximately 15\% larger than the values obtained from
direct SED fitting.
Comparison of our results with the independent near-IR extinction
measurement of the B68 column density profile by \cite{alves2001}
suggests that the dust opacity law adopted by the HGBS consortium,
with $\kappa_{300~\mu \rm m} = 0.1\, \rm{cm}^{2}$ per g (of gas $+$
dust) at $\lambda = 300$ \micron\ and $\beta = 2$, is 
accurate to better than (and possibly overestimated by) 50\% 
in the 160--500 \micron\ range for sources of
(column) densities comparable to B68 and L1689B. 
Our adopted opacity value is within $\sim$20\% of the value \footnote{ \citealp{suutarinen2013}
quote $\kappa_{250 \mu \rm m}$ =0.08 cm$^2$ g$^{-1}$ with $\beta$=2 and
argue that this value is underestimated by 40\% inside the core due to
temperature variations along the LOS.} 
 obtained by \cite{suutarinen2013}  for the dust 
inside a core of similar column density to the ones considered here. 
 Assuming that the
weak trend between submillimeter dust opacity and column density
($\kappa_{\lambda} \propto \, $\nhtwo$^{0.28}$) found by
\cite{roy2013} at $ A_V \la 10$ also holds at higher $ A_V $, we argue
that the HGBS dust opacity law may remain valid to within 50\%
accuracy in the whole range of H$_2$ column densities between $\sim 3
\times 10^{21}\, {\rm cm}^{-2} $ and $\sim 10^{23}\, {\rm cm}^{-2} $.

Since the Abel transformation technique can help us correct the effect
of LOS temperature variations on the derivation of total masses, we
conclude that the $Herschel$ data of the HGBS project hold the promise
of delivering core masses to better than a factor of 1.5 to 2 accuracy, at
least for spatially-resolved cores.

\begin{acknowledgements}
This work has benefited from the support of CNES and 
the European Research Council under 
the European Union's Seventh Framework Programme (Fp7/2007-2013 -- ERC Grant Agreement no. 291294).
P.P acknowledges funding by the Funda\c{c}\~ao para a Ci\^encia e a Tecnologia
(Portugal).  SPIRE has been developed by a consortium of institutes
led by Cardiff Univ. (UK) and including Univ. Lethbridge (Canada);
NAOC (China); CEA, LAM (France); IFSI, Univ. Padua (Italy); IAC
(Spain); Stockholm Observatory (Sweden); Imperial College London, RAL,
UCL-MSSL, UKATC, Univ. Sussex (UK); Caltech, JPL, NHSC, Univ. Colorado
(USA). This development has been supported by national funding
agencies: CSA (Canada); NAOC (China); CEA, CNES, CNRS (France); ASI
(Italy); MCINN (Spain); SNSB (Sweden); STFC (UK); and NASA (USA). PACS
has been developed by a consortium of institutes led by MPE (Germany)
and including UVIE (Austria); KUL, CSL, IMEC (Belgium); CEA, OAMP
(France); MPIA (Germany); IFSI, OAP/AOT, OAA/CAISMI, LENS, SISSA
(Italy); IAC (Spain). This development has been supported by the
funding agencies BMVIT (Austria), ESA-PRODEX (Belgium), CEA/CNES
(France), DLR (Germany), ASI (Italy), and CICT/MCT (Spain).
\end{acknowledgements}


\begin{appendix}

\section{Surface brightness profiles and outer radius of B68}\label{appen:profile}

\begin{figure}
  \begin{center}
  \resizebox{\hsize}{!}{\includegraphics{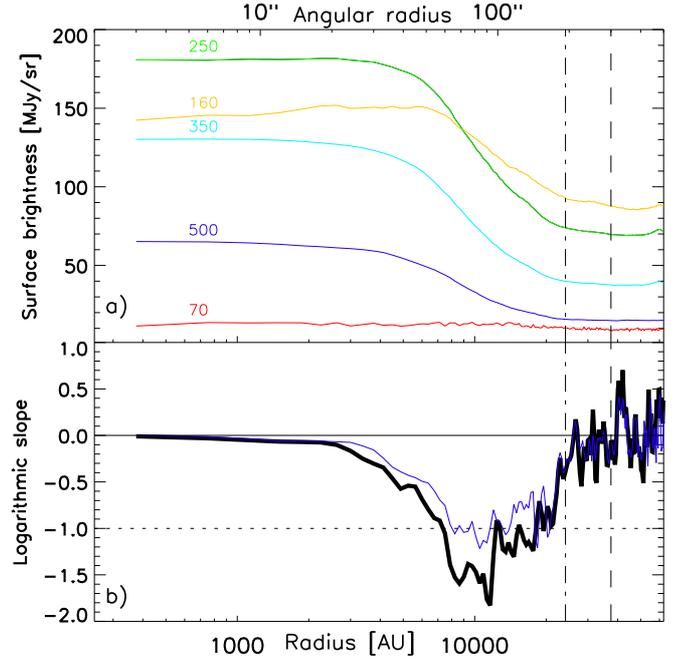}}
  \end{center}
  \caption{{\bf a)} Circularly-averaged radial surface brightness
    profiles of B68 at 70, 160, 250, 350, 500~$\mu$m derived from
    \emph{Herschel}/SPIRE and PACS data (after adding Planck offsets).
         {\bf b)} Logarithmic slopes of the circularly-averaged column
         density profile (black solid curve) and 500~$\mu$m intensity
         profile (blue solid curve) of B68 as a function of radius.
         The horizontal dotted line is the logarithmic slope $s = -1$
         expected for the column density profile of 
         a core with a $\rho \propto r^{-2}$ density
         profile.  The vertical dashed line marks the integration
         upper bound adopted when integrating the right-hand side of
         Eq.~\ref{eq:b} to perform the Abel reconstruction of the
         density and temperature profiles (see
         Sect.~\ref{sec:abel}). The vertical dot-dashed line shows 
         the radius of the  footprint automatically derived for B68 by the
         \emph{getsources} source-finding algorithm.  }
  \label{fig:AA1}
\end{figure}

The top panel of Fig.~\ref{fig:AA1} shows the circularly-averaged
intensity profiles of the B68 core at SPIRE and PACS wavelengths which
were used to construct the column density and temperature profiles
shown in Fig.~\ref{fig:B68-profile}.  The bottom panel of
Fig.~\ref{fig:AA1} shows the logarithmic slope of the column density
profile (black line), defined as $s \equiv d\,{\rm ln}\, {N_{\rm
    H_2}}/d\,{\rm ln}\,r $ (dimensionless), as a function of radius.
The measured logarithmic slope is $s = 0$ near core center due
to flat inner density profile and the finite resolution of the observations. 
The logarithmic slope profile reaches a minimum value  $s_{\min}
\sim-1.5$ at $r \sim 10^4$~AU, and goes back to $s \sim 0$ at the
outer boundary where the core merges with the slowly-varying
background.  For comparison, a spherical core with outer density
profile $\rho \propto r^{-2}$ would have $s = -1$ at large radii. In
the same plot we show the logarithmic slope of the 500~\micron\ 
surface brightness profile (blue line)  which has a shallower slope 
due to the additional effect of the positive outward temperature gradient.
Inspection of the intensity and slope profiles shown in
Fig.~\ref{fig:AA1} allowed us to select an appropriate upper
integration radius in the right-hand side of Eq.~(\ref{eq:b}) when 
reconstructing the density and temperature structure of B68 (see
Sect.~\ref{sec:abel}).  Since $\frac{dI_\nu}{dp} = s \times
\frac{I_\nu}{p} \approx 0 $ beyond a radius $\sim 25000$~AU, larger
radii do not contribute to the integral of Eq.~(\ref{eq:b}). In
practice, we adopted an upper integration radius  $R_{\rm up} =
37500$~AU for B68, as shown by the dashed vertical line in Fig.~\ref{fig:AA1}. 
The reconstruction results, however, are insensitive to the precise choice
of $R_{\rm up} $ as long as $R_{\rm up} \ga 25000$~AU.
The adjacent dot-dashed vertical line shows the effective radius of $\sim 27000$~AU
(or $\sim 200\arcsec $) derived by the \emph{getsources} source-finding 
algorithm for the `footprint' of B68. 
The \emph{getsources} algorithm \citep{mensch2012} is the source extraction method
used by the HGBS consortium to produce the first-generation catalogs of dense cores 
found by  \emph{Herschel} in the regions covered by the HGBS survey. 
The footprint of a core corresponds to the area just outside of which \emph{getsources} 
estimates the local background emission and 
over which it integrates the background-subtracted emission to derive 
the total flux densities of the core.  
In the case of B68, the results automatically derived by \emph{getsources} 
are in excellent  agreement with those obtained through a detailed radial profile 
analysis (cf. Fig.~\ref{fig:AA1}).

\section{Tests of the Abel inversion method using simple models}\label{appen:sim}
\subsection{Spherically symmetric core model}\label{A:sph}

\begin{figure}
  \begin{center}
  \resizebox{\hsize}{!}{\includegraphics{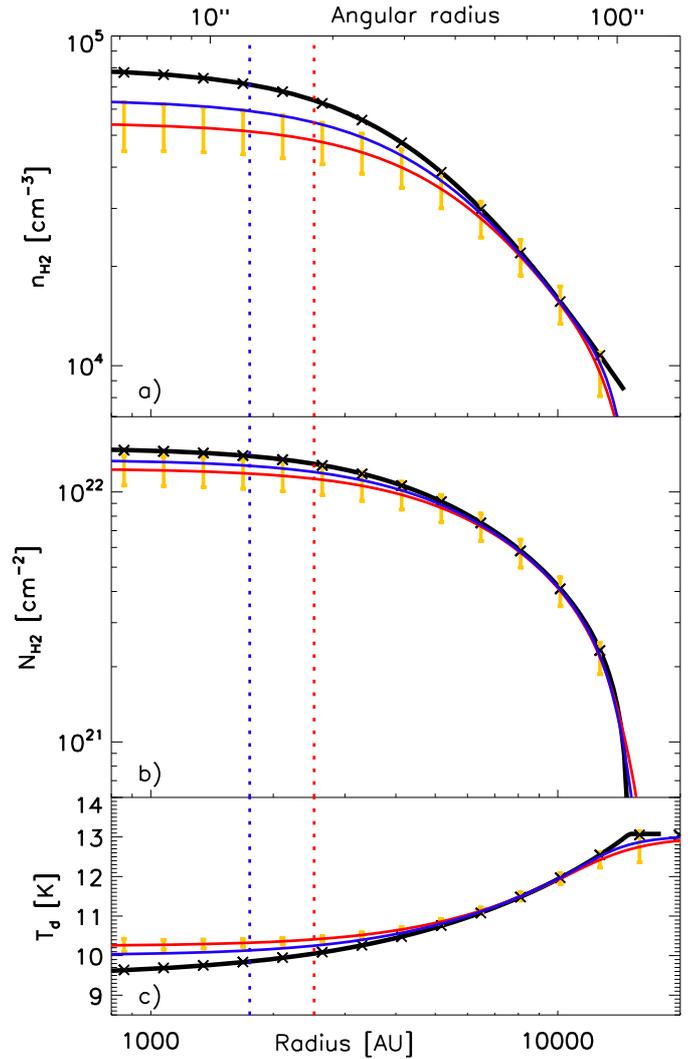}}
  \end{center}
  \caption{
Comparison between intrinsic (black curves) and reconstructed (red and
blue curves) volume density {\bf (a)}, column density {\bf (b)}, and
dust temperature {\bf (c)} profiles for a spherically symmetric core model with
a Plummer-like density distribution (see Eq.~\ref{eq:sphere} and text
for model parameters).  The cross symbols show the results obtained by
applying the Abel reconstruction scheme on the synthetic
160--500~\micron\ intensity profiles with ``infinite'' resolution.
The red and blue curves show the reconstruction results obtained from
synthetic emission maps smoothed to HPBW resolutions of
36\farcs3 and 24\farcs9, respectively, corresponding to the
resolutions of \emph{Herschel} 500~\micron\ and
350~\micron\ observations.  Note the good agreement between the
reconstructed profiles and the intrinsic profiles beyond the the half
power beam radius of 36\farcs3/2 (500~\micron\ resolution) or
24\farcs9/2 (350~\micron\ resolution), marked by the red and blue
vertical dotted lines, respectively. The error bars in each panel
  show the 1-$\sigma$ uncertainties due to random noise and calibration errors.}
  \label{fig:AA2}
\end{figure}

In order to test the performance level of our Abel inversion scheme
and quantify the robustness of the reconstruction, we applied the
algorithm described in Sect.~\ref{sec:abel} to synthetic images
corresponding to model starless cores of known density and temperature
distributions.  First, we considered a spherically-symmetric core
model with a Plummer-type density distribution for $r \le R_{\rm out}$,
\begin{equation}
\rho(r)= \frac{\rho_{\rm c}} {1+(r/R_{\rm flat})^2}   ,
\label{eq:sphere}
\end{equation}
parameterized by physical parameters approximately similar to the
derived properties of B68 (see Sect.~\ref{sec:B68}): central H$_2$
number density $n_{\rm c} \equiv \rho_{\rm c} /\mu_{\rm H_2}{\rm m_H}
=8\times10^{4}$ cm$^{-3}$; flat inner radius \Rflat\ $= 5 \times10^3$
AU, and outer radius $R_{\rm out} = 1.5\times10^{4}$ AU.  
The surface density profile of such a model core has an analytical form:
\begin{equation}
\Sigma(p) =\frac{2\rho_{\rm c}R_{\rm flat}}{\left(1+p^2/R^2_{\rm flat}\right)^{1/2}} \times  {\rm tan^{-1}}
\left( \frac{(R^2_{\rm out} -p^2)^{1/2}}{(R^2_{\rm flat} +p^2)^{1/2}}    \right), 
\end{equation} 
where $p$ represents the impact parameter from core center in the plane of the sky, 
and \nhtwo$(p)$=$\Sigma(p)/\mu_{\rm H_2}m_{\rm H}$ is the H$_2$ column density profile.
The intrinsic density profile of the model is shown as a black solid curve
in Fig.~\ref{fig:AA2}a, and the corresponding column density profile
as a black solid curve in Fig.~\ref{fig:AA2}b.
The synthetic dust temperature profile is shown as a black solid curve in
Fig.~\ref{fig:AA2}c and was obtained for a solar-neighborhood ISRF
($G_0 = 1$) using an analytic approximation formula reproducing  a grid of
spherically symmetric models performed with the dust radiative
transfer code MODUST (Bouwman et al. 2013, \emph{in preparation} --
see \citealp{bouwman2001} and \citealp{andre2003}).

A set of synthetic emission maps was created by line-of-sight
integration of this model core at all $Herschel$ wavelengths assuming
optically thin dust emission
(see Eq.~\ref{eq:a}) and the same dust
opacity law as given in Sect.~\ref{sec:sedfit}.  The density and
temperature profiles of the model core were then reconstructed as
described in Sect.~\ref{sec:abel} from the circularly-averaged radial
intensity profiles of the model emission.  The cross symbols overlaid
on the model density, column density, and temperature profiles in
Figs. \ref{fig:AA2}a,b,c show the Abel-reconstructed profiles that
would be obtained with ``infinite'' angular resolution (and in the
absence of noise). It can be seen in Fig. \ref{fig:AA2} that, in this
case, the reconstruction is perfect, demonstrating the validity of our
Abel-inversion code.  The overplotted red curves in
Figs.~\ref{fig:AA2}a,b,c show the reconstructed volume density, column
density, and temperature profiles resulting from the Abel-inversion
method after convolution of the model images to a common resolution of
36\farcs3\, corresponding to the \emph{Herschel} resolution at
500 \micron .  Likewise, the overplotted blue curves in
Figs.~\ref{fig:AA2}a,b,c show the results obtained at a resolution of
24\farcs9, using the synthetic data convolved to the \emph{Herschel}
resolution at 350-\micron\ and ignoring the 500-\micron\ data.  It can
be seen that the profiles reconstructed at the \emph{Herschel}
resolution remain in excellent (1\%) agreement with the intrinsic profiles
in the outer part of the core.  Although the
reconstruction becomes somewhat inaccurate below the \emph{Herschel}
resolution limit (marked by vertical dotted lines in
Figs.~\ref{fig:AA2}), the reconstructed column density and temperature
profiles still agree with the corresponding intrinsic profiles to
within 20\% and 9\%, respectively, at 500-\micron\ resolution.  The
accuracy of the results at small radii improves to 11\% and 5\% when the
reconstruction is performed at 350-\micron\ resolution (although in
the presence of noise with real data, the statistical measurement
uncertainties are somewhat larger at 350-\micron\ resolution).
At both resolutions, the Abel-reconstructed temperature and column density profiles
coincide  within 1\% with the corresponding intrinsic  profiles {\it convolved} 
with the effective beam resolution.
The reconstructed central temperature and column density   thus provide excellent estimates 
of the beam-averaged central temperature and column density in the model. 
The total mass estimated by integrating the reconstructed column
density profile agrees with the model mass to better than 0.1\% even
at 500-\micron\ resolution.

We also assessed the contribution of background fluctuations
and calibration errors to the uncertainties in the derived parameters (\nhtwo and \Td).
To do so, we considered 500 realizations of synthetic skies including a
random Gaussian noise component\footnote{The level of noise fluctuations 
($\sigma_{\nu}$)  was chosen so that the peak 
signal-to-noise  $I_{\nu}^{\rm peak}/\sigma_{\nu}$  value at each wavelength was consistent with the
corresponding B68 surface brightness image.}, $\sigma_{\nu}$,  and a random multiplicative calibration factor, ($1+g$):
\begin{equation}
I_{\nu}^{\rm sim}(x,y)= [I_{\nu}^{\rm model}(x,y) +\sigma_{\nu}](1+g),
\end{equation}
where, $g$ is a Gaussian random number with mean zero and standard deviation of 10\% and 15\% at SPIRE and PACS wavelengths,
respectively.
(We assumed 100\% correlated calibration  errors at  
SPIRE wavelengths and an independent calibration error in the PACS 160~\micron\ band.)
The net uncertainties in $n_{\rm H_2}$, \nhtwo, and \Td\  were estimated to be 15\%, 12\%, and
5\%, respectively.
The resulting 1-$\sigma$ errors in the derived parameters  are displayed in Fig.\ref{fig:AA2}.

\begin{figure}[!!!b]
   \centering
 \begin{minipage}{1.1\linewidth}
   \begin{minipage}{.85\linewidth}
   \resizebox{1\hsize}{!}{\includegraphics[angle=270]{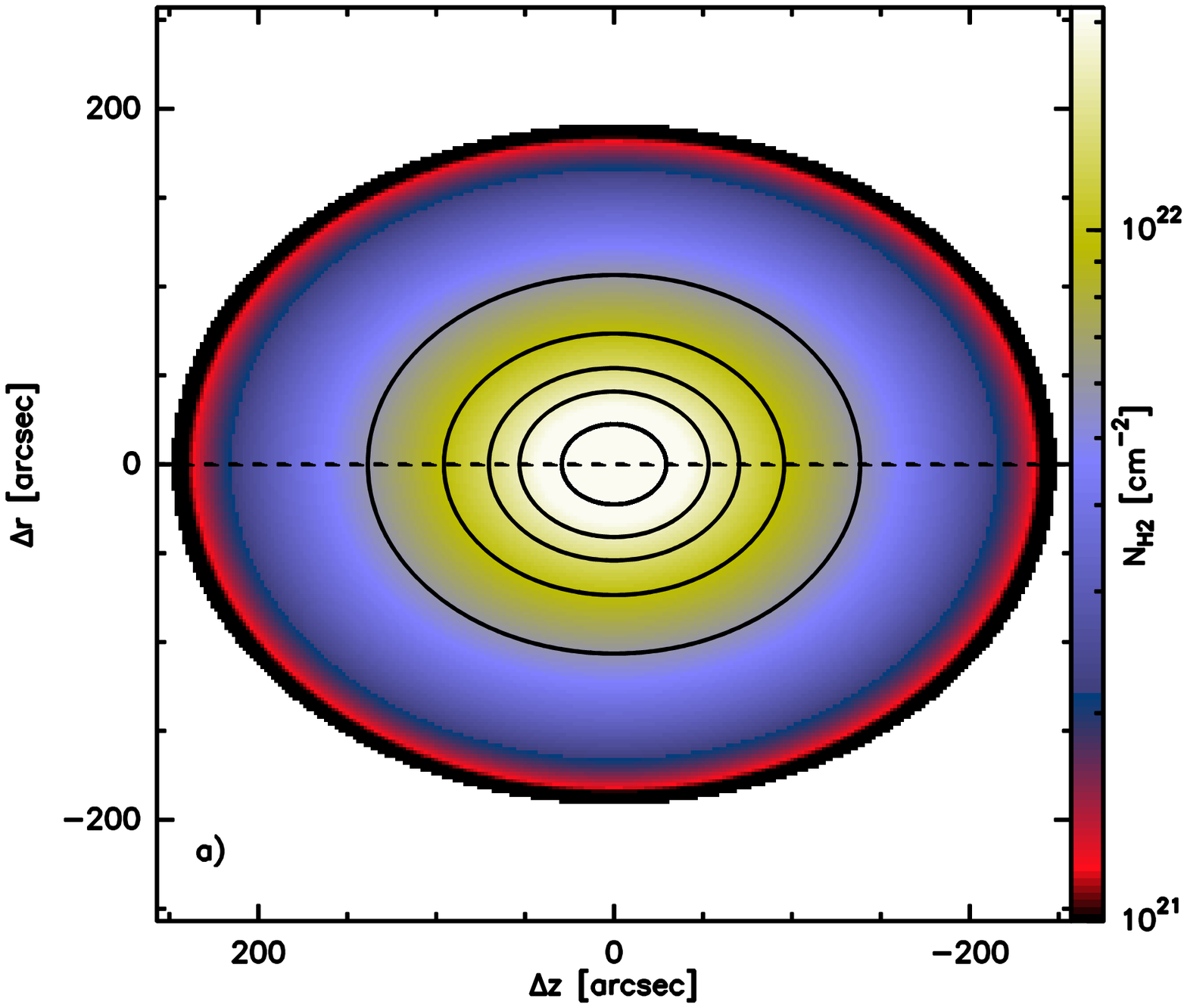}}
   \vspace{4mm}
   \end{minipage}
   \begin{minipage}{.8\linewidth}
   \resizebox{1\hsize}{!}{\includegraphics[angle=0]{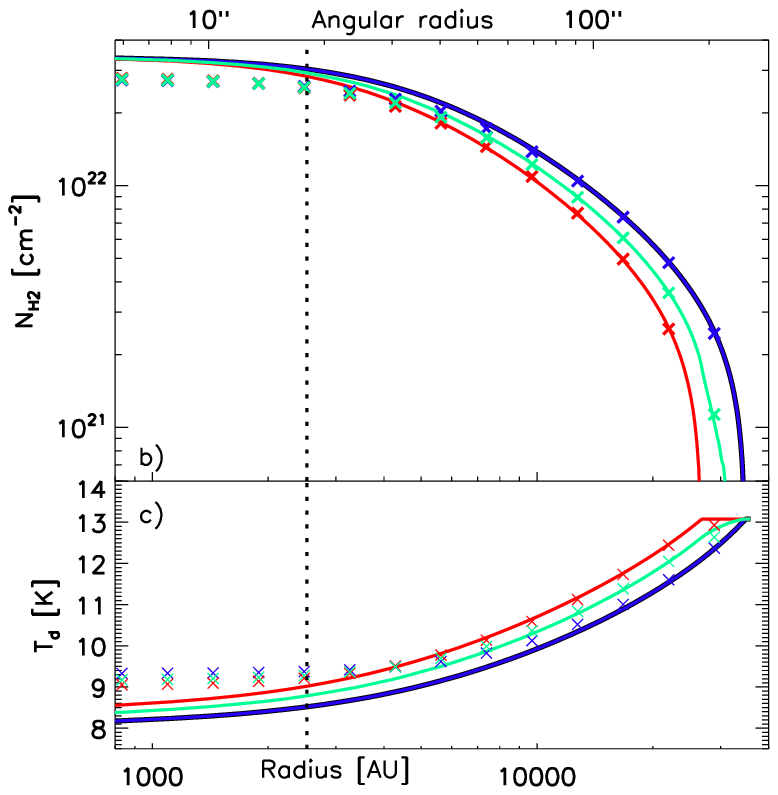}}
   \end{minipage}
 \end{minipage}
   \caption{ {\bf (a)} Synthetic column density image of a prolate ellipsoidal 
core model with an aspect ratio of 1.3 and a Plummer-like density distribution 
(see Eq.~(\ref{eq:spheroid}) and the text for model parameters).
The horizontal dotted line shows the axis of symmetry in the plane of sky 
and the contour levels are same as in Fig.~\ref{fig:L1689B-nhT}. 
{\bf (b)} Comparison between the intrinsic (solid curves) and the reconstructed (crosses) 
column density profiles of the model.
{\bf (c)} Comparison between the intrinsic (solid curves) and the reconstructed (crosses) 
dust temperature profiles of the model.
   The blue and red curves
    represent the intrinsic radial profiles 
    along and
    perpendicular to the axis of symmetry, respectively.  The green
   curves represent the intrinsic circularly-averaged radial profiles.
    The blue and red crosses display the results of the
    Abel-inversion method applied to the synthetic
    160--500~\micron\ intensity profiles of the model convolved to
    36\farcs3 resolution and taken along the major and minor axes,
    respectively.  The green crosses show the reconstruction results
    using the circularly-averaged intensity profiles of the model as
    inputs.  Note the good agreement between the reconstructed
    profiles and the intrinsic profiles beyond the half power beam
    radius of 36\farcs3/2, marked by the vertical dotted line.
}
     \label{fig:AA3}%
    \end{figure}

\subsection{Ellipsoidal core model}\label{A:ellip}

As real cores such as L1689B are often elongated and thus not strictly
spherically symmetric (see Sect.~\ref{sec:L1689B}), we also tested the
reliability of our Abel inversion scheme using a simple non-spherical
model with an ellipsoidal Plummer-like density distribution for $r \le
R_{\rm out}$ and $z \le Z_{\rm out}$, with cylindrical symmetry about
the $z$ axis (assumed to lie in the plane of the sky):
\begin{equation}
\rho(r,z)=\frac{\rho_c}{1+(r/R_{\rm flat})^2+(z/Z_{\rm flat})^2   },
\label{eq:spheroid}
\end{equation}
where $R_{\rm flat}$ and $Z_{\rm flat}$ are the radii of the flat
inner core region perpendicular and parallel to the $z$ axis of
symmetry (see Fig.~\ref{fig:AA3}a), respectively.  
We considered both the prolate ($Z_{\rm
  flat}$ $>$ $R_{\rm flat}$) and the oblate ($Z_{\rm flat}$ $<$
$R_{\rm flat}$) configuration, but are primarily describing the
prolate case here as it is more likely for cores embedded within
filaments such as L1689B (see Fig.~\ref{fig:L1689B-nhT}).  The
synthetic temperature distribution was also assumed to be
cylindrically symmetric about the $z$ axis and was constructed using
the same grid of MODUST radiative transfer models as in
Sect.~\ref{A:sph}. The synthetic temperature profiles along both the
$z$ axis and the radial ($r$) direction are shown in Fig.~\ref{fig:AA3}.
For direct comparison with L1689B (see Fig.~\ref{fig:L1689B-nhT} and
Fig.~\ref{fig:AA3}a), we adopted physical parameters
approximately consistent with the observed characteristics of the
L1689B core (see Sect.~\ref{sec:L1689B}): central H$_2$ number density
$n_{\rm c} = 2\times10^{5}$ cm$^{-3}$; aspect ratio $Z_{\rm
  flat}/R_{\rm flat} = Z_{\rm out}/R_{\rm out} = 1.3$; flat inner
radius along the minor axis $R_{\rm flat} = 4000$~AU; outer radius
along the minor axis $R_{\rm out} = 6.7 \times R_{\rm flat} =
26800$~AU.

Because of the lack of spherical symmetry we applied our Abel
reconstruction scheme to three sets of intensity profiles: 1) the
profiles measured along the major axis of the model (intrinsic
profiles shown as blue curves and results as blue crosses in
Fig.~\ref{fig:AA3}); 2) the profiles measured along the minor axis of
the model (intrinsic profiles shown as red curves and results as red
crosses in Fig.~\ref{fig:AA3}); and 3) circularly-averaged intensity
profiles (intrinsic profiles shown as green curves and results as
green crosses in Fig.~\ref{fig:AA3}).  Here, again, it can be seen
that the reconstruction results are very satisfactory (2\% agreement)
beyond the beam radius (marked by the vertical dotted line in 
Fig.~\ref{fig:AA3}).  The reconstruction performed perpendicular to
the axis of symmetry, i.e., along the minor axis for a prolate core, is
more accurate (1\%) than the reconstruction performed along the axis of
symmetry (4\%). In particular, the best estimate of the central
dust temperature is obtained from the reconstruction performed along
the minor axis. The reconstruction along the major axis nevertheless
provides better estimates of the column density and temperature at
large radii along the major axis.  The central column density
reconstructed at 500~\micron\ resolution slightly underestimates, but
still agrees to within 20\% with, the true column density at core
center.  The best estimate of the total core mass, obtained by using
the results of the reconstruction performed on the circularly-averaged
intensity profiles, agrees to better than 4\% with the model core
mass. Even for a more elongated core model with an aspect ratio of 2 
(instead of 1.3), the reconstructed core mass still agrees with the model
mass to within 5\%.

We also performed similar simulations for an oblate core model observed edge-on.  
The accuracy of the reconstruction results was found to be essentially the same 
as for the prolate case. Again, the reconstruction performed perpendicular to
the axis of symmetry, i.e., along the major axis in this case, was found to be 
more accurate than the reconstruction performed along the axis of symmetry.
The best estimate of the total core mass was again obtained from reconstructing 
the circularly-averaged intensity profiles.

\end{appendix}

\end{document}